\documentclass{svproc}
\usepackage{url}

\usepackage{graphicx}
\usepackage{hyperref}
\usepackage{url}

\usepackage{times}
\usepackage{mathptmx}

\begin{document}
\mainmatter  
\title{Heavy-flavour measurements in p-Pb collisions with ALICE at the LHC}
\author{Jitendra Kumar, for the ALICE collaboration}
\institute{Indian Institute of Technology Bombay, Mumbai, India,\\ \email{jitendra.kumar@cern.ch}}
\maketitle    

\begin{abstract}
The measurements of open heavy-flavours, i.e. D mesons at central rapidity and leptons from charm and beauty decays at central and forward rapidity was studied in p-Pb collisions at $\sqrt{s_{{\rm{NN}}}}$ = 5.02 TeV using the ALICE detector. The results are presented and compared to model predictions including cold nuclear matter effects. 
\end{abstract}

\section{Introduction}
Heavy quarks (charm and beauty) due to their large masses are predominantly produced in hard-scattering processes in the initial phase of hadronic collisions. Therefore, they are excellent probes to study the properties of the Quark-Gluon Plasma created in relativistic heavy-ion collisions. The measurement of their production in p-Pb collisions is important to disentangle the hot nuclear matter effects present in heavy-ion collisions from cold nuclear matter (CNM) effects, such as transverse momentum broadening, nuclear modification of the parton distribution functions, initial-state multiple scatterings and energy loss. These effects can be investigated by measuring the nuclear modification factor $R_{{\rm{pPb}}}$, defined as the ratio of particle cross section d$\sigma$/d$p_{\rm{T}}$ measured in p-Pb collisions to that measured in pp collisions scaled by the atomic mass number of Pb nuclei. In the absence of CNM effects $R_{{\rm{pPb}}}$ is expected to be unity. The $R_{{\rm{pPb}}}$ of D mesons and leptons from charm and beauty hadron decays at central and forward rapidities was studied in p-Pb collisions at $\sqrt{s_{{\rm{NN}}}}$ = 5.02 TeV with ALICE.


\section{Analysis details} 
Prompt D mesons and their charge conjugates are reconstructed via their hadronic decay channels: $\rm{D}^{0} \rightarrow \rm{K}^{+}\pi^{-}$, $\rm{D}^{+} \rightarrow \rm{K}^{-}\pi^{+}\pi^{+} $ and $\rm{D}^{*+} \rightarrow \rm{D}^{0}\pi^{+} $ \cite{alice2}. The extraction of the signal is based on an invariant mass analysis of reconstructed decay vertices displaced from the primary vertex by few hundred microns. The necessary spatial resolution on the track position is guaranteed by the Inner Tracking System (ITS) and the Time Projection Chamber (TPC) covering a pseudorapidity region $|\eta| < 0.8$. Particle identification (PID) of the decay particle species is also exploiting using the measurement of specific energy loss (${\rm d}E/{\rm d}x$) in the TPC and of the time of flight with the Time-Of-Flight (TOF) detector. Kaons and pions are identified up to {\normalfont\textit{p}$_{\mathrm{T}}$ = 2 GeV/$c$. The electrons from heavy-flavour (HF) hadron decays are identified using ITS, TPC and TOF detectors in the range 0.5 $<$ $p_{\rm{T}}$ $<$ 6 GeV/$c$ and using the TPC and the Electromagnetic Calorimeter (EMCal) for $p_{\rm{T}}$ $>$ 6 GeV/$c$ \cite{alice4}. The background from $\pi^{0}$ and $\eta$ Dalitz decay and from photon conversions is subtracted via the invariant mass method, and the hadron contamination decays is statistically subtracted \cite{alice4}. The muons from heavy-flavour hadron decays are measured with the muon spectrometer in pseudorapidity range, 2.5 $<$ $y_{\rm{lab}}$ $<$ 4 \cite{alice5} (additional information in \cite{alice1}). The background from $\pi$ and K decays is subtracted using a data-tuned Monte Carlo cocktail.

\begin{figure}[h!]
\centering
\vspace{-0.25cm}
\includegraphics[scale=.265]{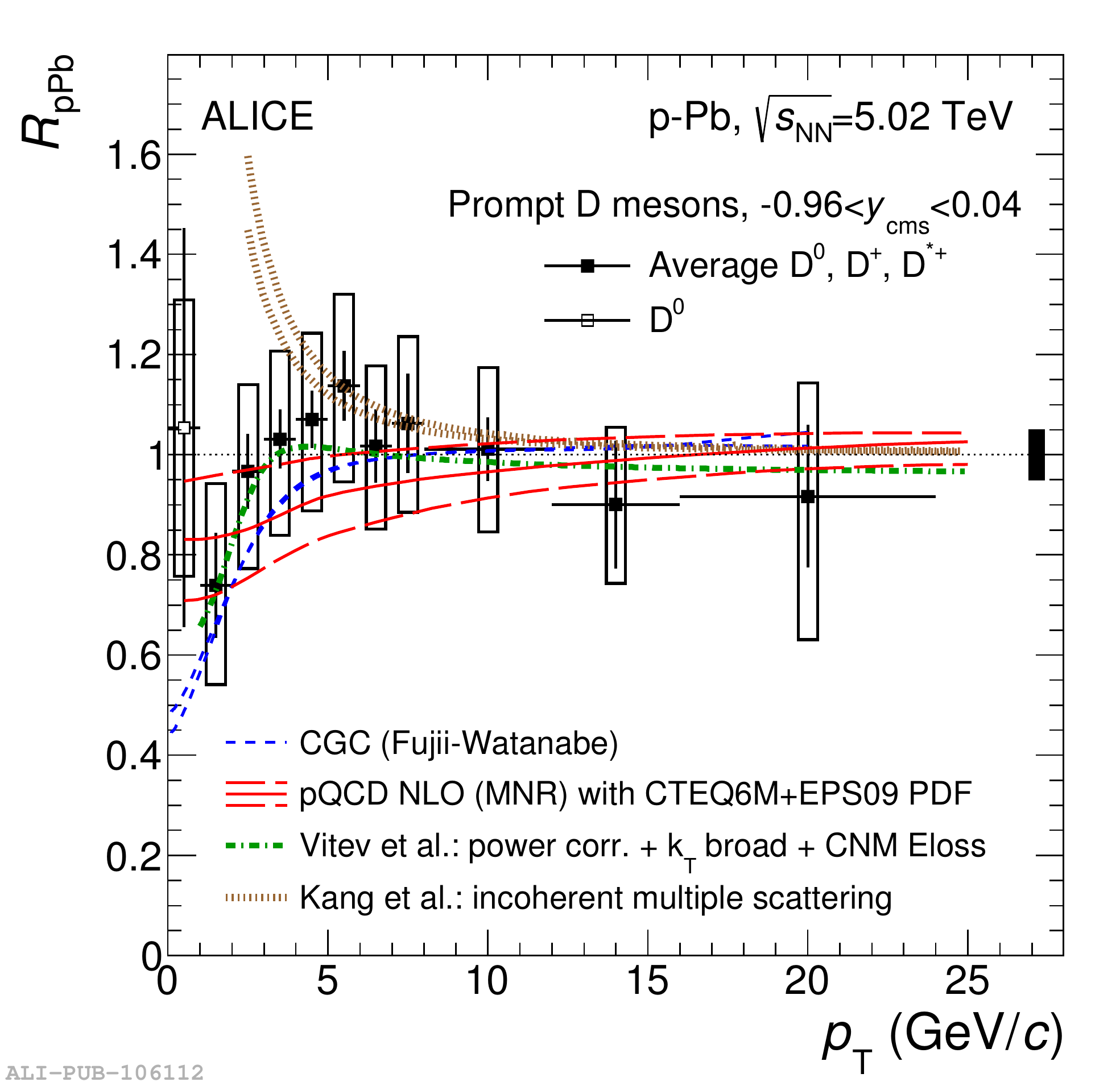}
\includegraphics[scale=.265]{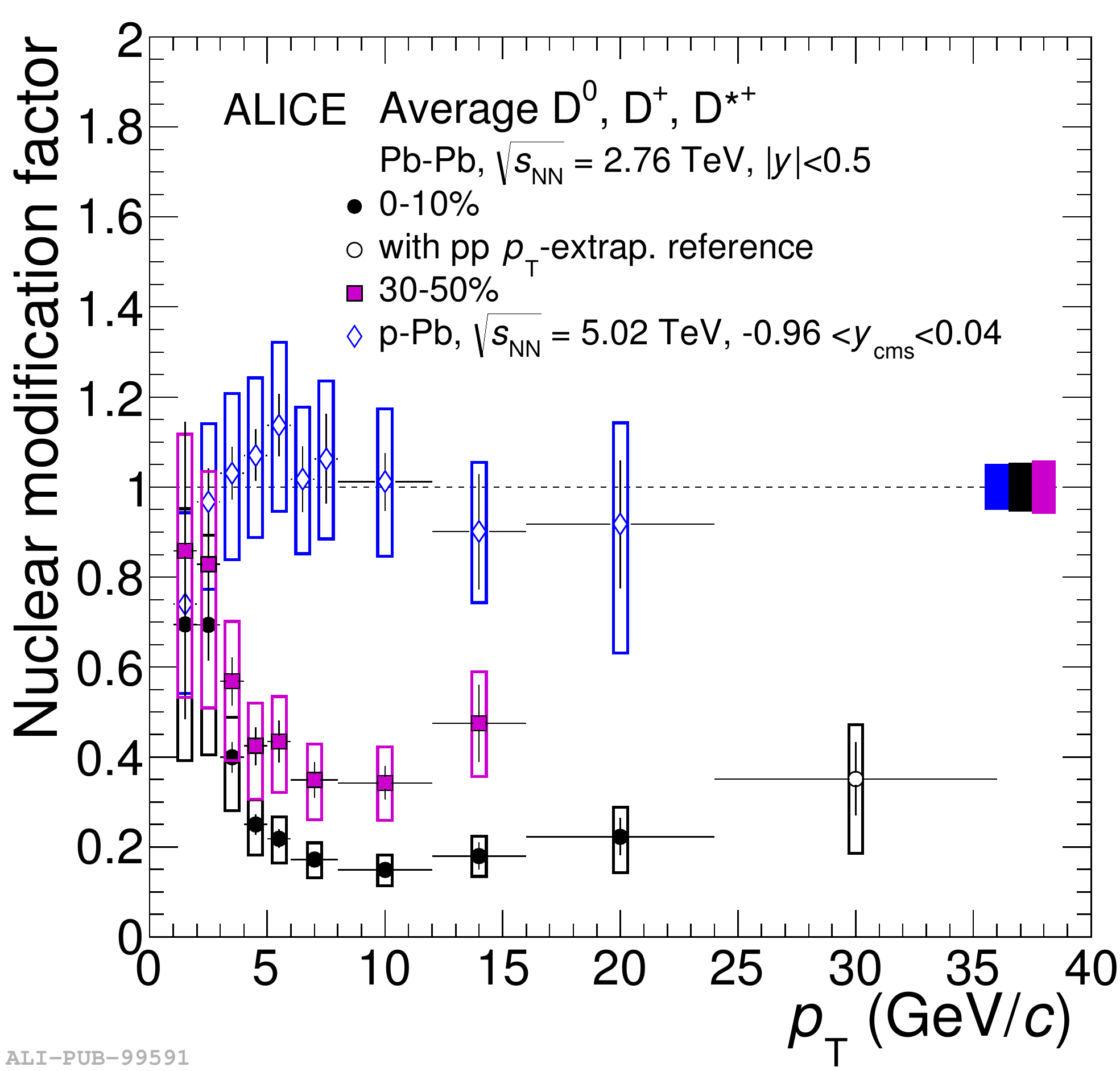}
\vspace{-0.20cm}
\caption{$R_{\rm{pPb}}$ of D mesons (D$^{0}$, D$^{+}$, and D$^{*+}$ average) compared with models, including CNM effects (left) and comparison with the $R_{\rm{AA}}$ for central and semi-central Pb-Pb collisions at $\sqrt{s_{{\rm{NN}}}}$ = 2.76 TeV (right) \cite{alice3}.}
\label{fig:RaaD}
\end{figure}

\vspace{-0.20cm}
\section{Results} 
The $R_{\rm{pPb}}$ of prompt D mesons (D$^{0}$, D$^{+}$ , and D$^{*+}$ average) is found compatible with unity, as shown in Figure~\ref{fig:RaaD} (left plot), and described by models which include CNM effects \cite{alice3}. The comparison to the nuclear modification factor in Pb-Pb collisions, $R_{\rm{AA}}$, is reported in Figure~\ref{fig:RaaD} (right plot) and highlights a strong suppression for $p_{\rm{T}}$ $>$  3 GeV/$c$ in central (0-10$\%$) and semi-central Pb-Pb collisions (30-50$\%$) \cite{alice0}. This comparison allows to conclude that the suppression observed in Pb--Pb collisions is due to final-state effects induced by the interaction of heavy quarks with the QGP produced in these collisions. The $R_{\rm{pPb}}$ of HF-hadron decay electrons shown in Figure~\ref{fig:RaaHFe} (left plot) is consistent with unity and also described by various models considering CNM effects \cite{alice4}. The impact parameter distributions of beauty decay electrons is expected to be broader than that of charm decay electrons due to the larger separation between the primary and decay vertices. Therefore, one can separate the contributions of charm and beauty production. The $R_{\rm{pPb}}$ of beauty decay electrons is shown in the right panel of Figure~\ref{fig:RaaHFe} \cite{alice6}. The results are similar and consistent with unity within the uncertainties.\\

\begin{figure}[ht!]
\centering
\includegraphics[scale=.265]{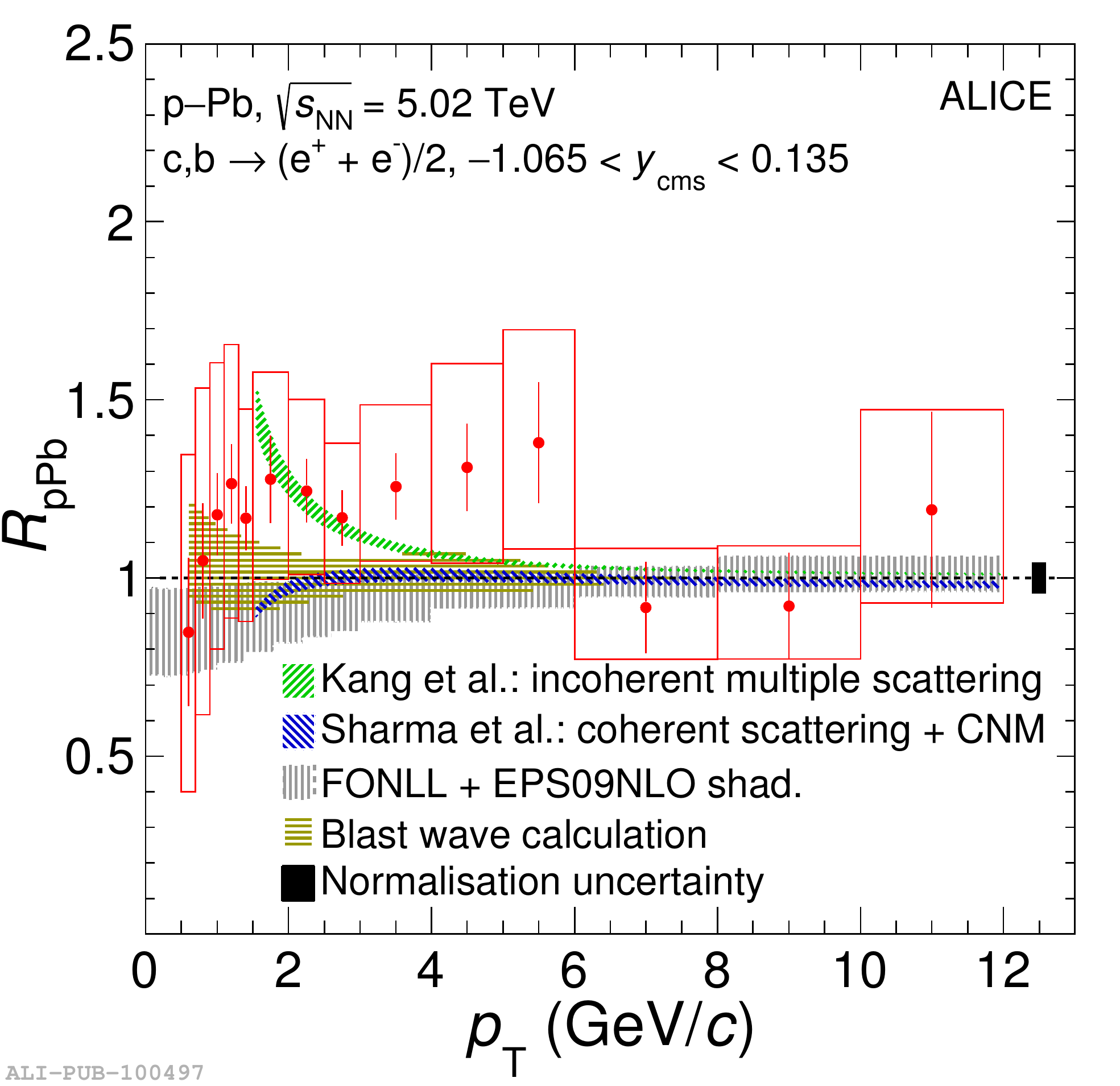}
\includegraphics[scale=.265]{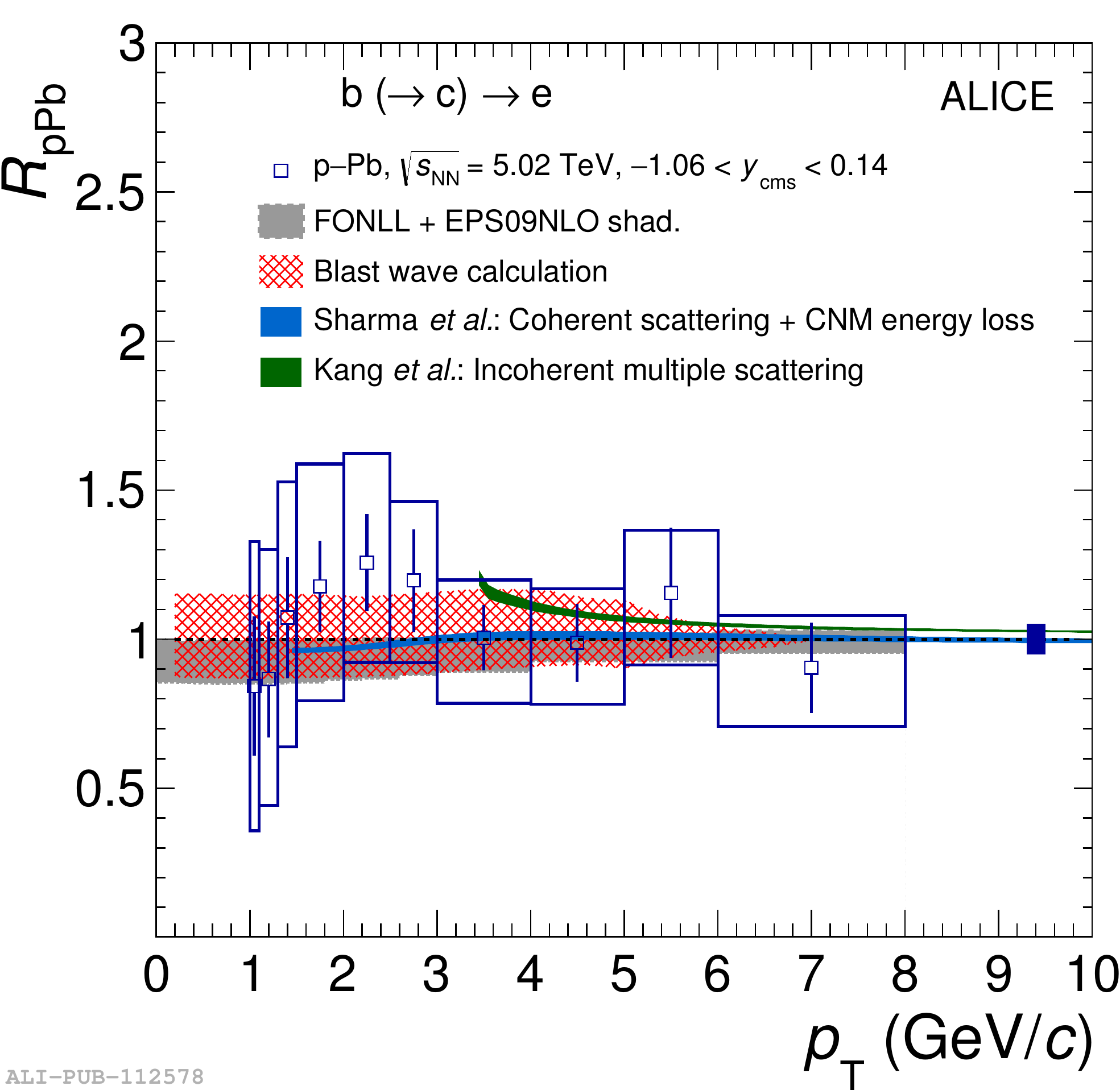}\\
\caption{$R_{\rm{pPb}}$ of inclusive electrons from HF-hadron decays (left plot, \cite{alice4}) and beauty-hadron decay electrons (right plot, \cite{alice6}) along with comparison to the models.}
\label{fig:RaaHFe}
\end{figure}

\noindent Figure~\ref{fig:RaaHFmu} shows the $R_{\rm{pPb}}$ of heavy-flavour hadron decay muons, which is also consistent with unity at both forward (2.03 $<$ $y_{\rm{cms}}$ $<$ 3.53, left panel) and backward (-4.46 $<$ $y_{\rm{cms}}$ $<$ -2.96, right panel) rapidities. However an enhancement is observed above unity at backward rapidity for 2 $<$ $p_{\rm{T}}$  $<$ 4 GeV/$c$ \cite{alice5}. The results in both rapidity ranges are described within uncertainties by model calculations that include CNM effects.

\begin{figure}[h!]
\centering
\includegraphics[scale=.300]{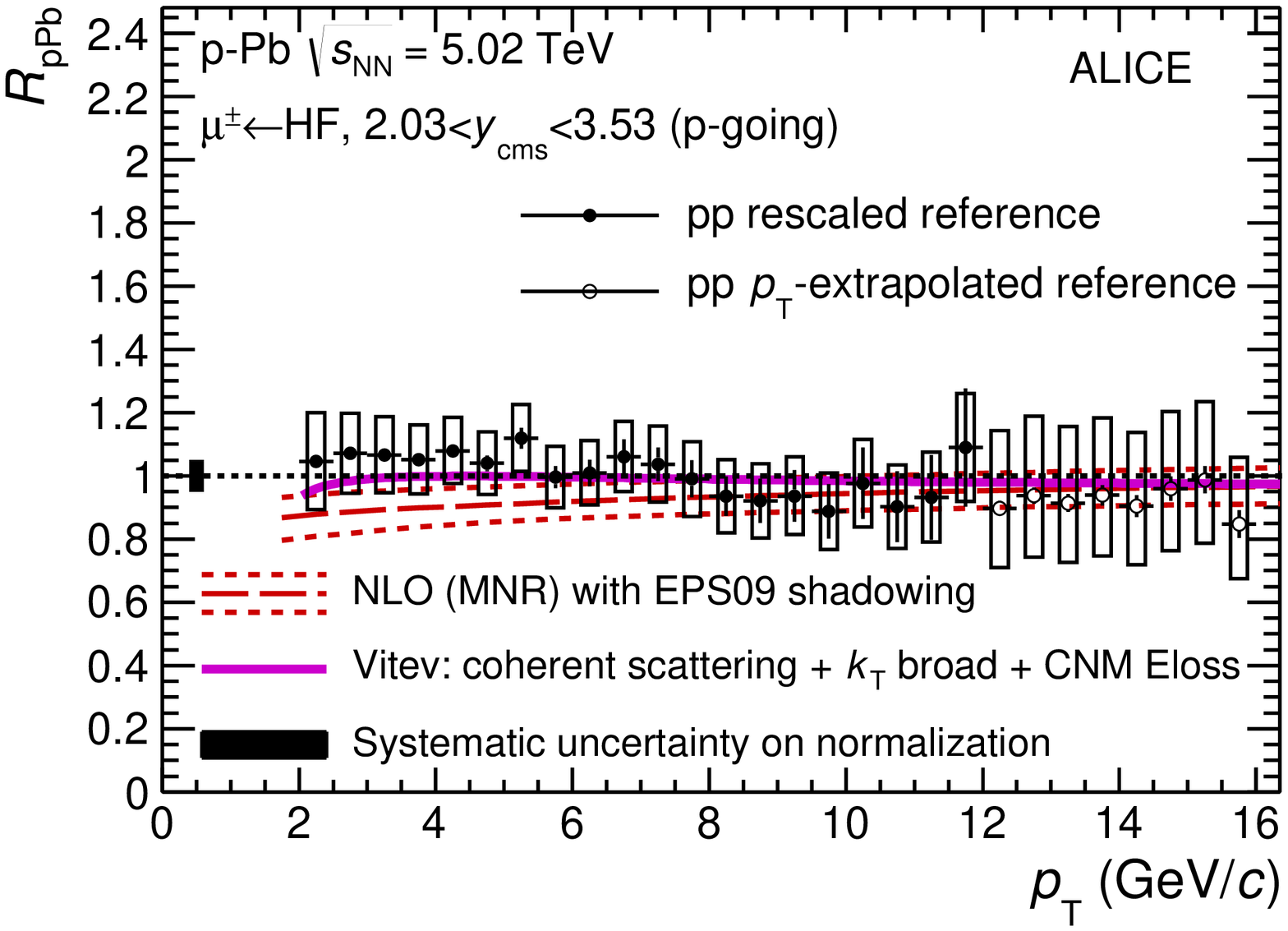}
\includegraphics[scale=.300]{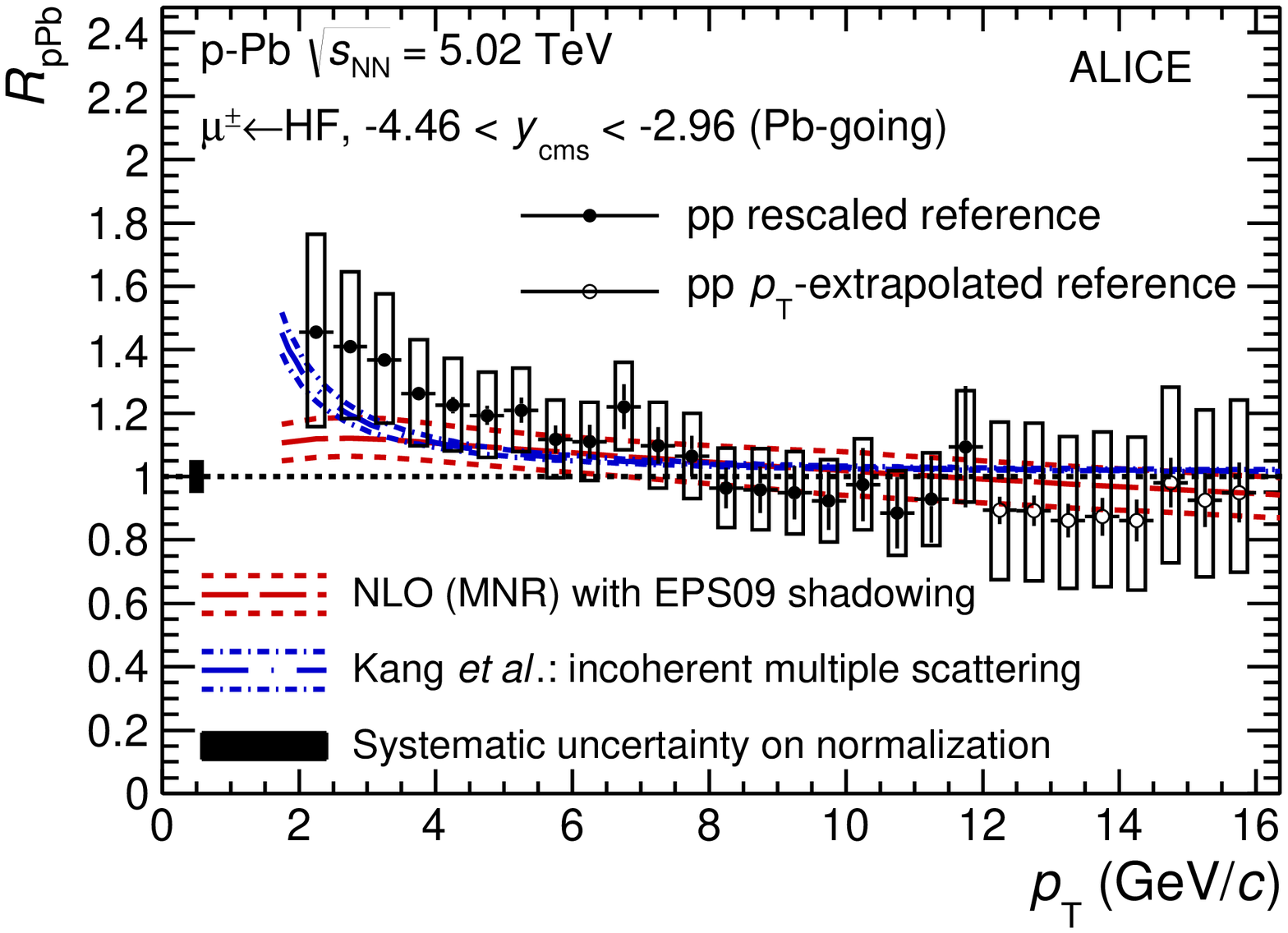}
\caption{$R_{\rm{pPb}}$ of muons from heavy-flavour hadron decays and comparison with models. Left (right):  forward (backward)
rapidity, 2.03 $<$ y$_{\rm{cms}}$ $<$ 3.53 (-4.46 $<$ y$_{\rm{cms}}$ $<$ -2.96) \cite{alice5}.}
\label{fig:RaaHFmu}
\end{figure}



\end{document}